\documentclass[amsmath,amssymb,showpacs,showkeys]{revtex4}
\usepackage{epsfig}
\usepackage{graphicx}

\begin{document}

\title{Symmetry and special relativity in Finsler spacetime with constant curvature }

\author{Xin Li$^{1,3}$}
\email{lixin@itp.ac.cn}
\author{Zhe Chang$^{2,3}$}
\email{changz@ihep.ac.cn}
\affiliation{${}^1$Institute of Theoretical Physics,
Chinese Academy of Sciences, 100190 Beijing, China\\
${}^2$Institute of High Energy Physics, Chinese Academy
of Sciences, 100049 Beijing, China\\
${}^3$Theoretical Physics Center for Science Facilities, Chinese Academy of Sciences}

\begin{abstract}
Within the framework of projective geometry, we investigate kinematics and symmetry in $(\alpha,\beta)$ spacetime-one special types of Finsler spacetime. The projectively flat $(\alpha,\beta)$ spacetime with constant flag curvature is divided into four types. The symmetry in type A-Riemann spacetime with constant sectional curvature is just the one in de Sitter special relativity. The symmetry in type B-locally Minkowski spacetime is just the one in very special relativity. It is found that type C-Funk spacetime and type D-scaled Berwald's metric spacetime both possess the Lorentz group as its isometric group. The geodesic equation, algebra and dispersion relation in the $(\alpha,\beta)$ spacetime are given. The corresponding invariant special relativity in the four types of $(\alpha,\beta)$ spacetime contain two parameters-the speed of light and a geometrical parameter which may relate to the new physical scale. They all reduce to Einstein's special relativity while the geometrical parameter vanishes.
\end{abstract}

\maketitle
\section{Introduction}
Lorentz Invariance (LI) is one of the foundations of the Standard
model of particle physics. Of course, it is very interesting to test
the fate of the LI both on experiments and theories. The theoretical approach of investigating the LI violation is studying the possible spacetime symmetry, and erecting some counterparts of special relativity. Recently, there are a few counterparts of special relativity. The first one is doubly special relativity (DSR)\cite{Amelino1,Amelino2,Amelino3}. In DSR, the Planck-scale effects have been taken into account by introducing an invariant Planckin parameter $\kappa$. Together with the speed of light $c$, DSR has two invariant parameters. The second one is very special relativity (VSR) \cite{Coleman1,Coleman2}. Coleman and Glashow have set up a perturbative framework for investigating possible departures of local quantum field theory from LI. The symmetry group of VSR is some certain subgroups of Poincare group, which contains the spacetime translations and proper subgroups of Lorentz transformations. The last is the de Sitter(dS)/anti de Sitter(AdS) invariant special relativity (dSSR) \cite{Look,Look1}. The dSSR suggests that the principle of relativity should be generalized to constant curvature spacetime with radius $R$ in Riemannian manifold.

In fact, the three kinds of modified special relativity share common ground. Historically, Snyder proposed a quantized spacetime model \cite{Snyder}. In his model, the spacetime coordinates were defined as translation generators of dS-algebra $\mathfrak{so}(1,4)$ and become noncommutative. It has already been pointed out in Ref. \cite{Guo} that there is a dual one-to-one correspondence between Snyder's quantized spacetime model as a DSR and the dSSR. Actually, the Plackin parameter $\kappa$ in DSR is related to the parameter $a$ in Snyder's model in addition to $c$. Furthermore, the dSSR can be regarded as a spacetime counterpart of Snyder's model.  VSR can be realized on a noncommutative Moyal plane with light-like noncommutativity \cite{Sheikh}. Thus, the three kinds of modified special relativity all have noncommutative realization.

On the other hand, these counterparts of special relativity have connections with Finsler geometry \cite{Bao}, which is a natural generalization of Riemannian geometry. The noncommutativity effects may be regarded as the deviation of Finsler spacetime from Riemann spacetime. Ref.\cite{Ghosh} gave a canonical description of DSR and showed that the DSR admits a modified dispersion relation (MDR) as well as noncommutative $\kappa$-Minkowskian phase space. Furthermore, Girelli {\it et al}.\cite{Girelli} showed that the MDR in DSR could be incorporated into the framework of Finsler geometry. As for VSR, Gibbons {\it et al}. have pointed out that general VSR is Finsler Geometry \cite{Gibbons}.

Therefore, It is reasonable to assume that these counterparts of special relativity may have a corporate origin in Finsler geommetry. In order to investigate the counterpart of special relativity in a systematic way, first, we should erect the inertial frames in Finsler spacetime. Second, we should investigate the symmetry in Finsler spacetime. The way of describing spacetime symmetry in a covariant language (the symmetry should not depend on any particular choice of
coordinate system) involves the concept of isometric transformations.
In fact, the symmetry of spacetime is described by the so called
isometric group. The generators of isometric group is directly
connected  with the Killing vectors\cite{Killing}. Actually, the
symmetry of deformed relativity has been studied by investigating
the Killing vectors\cite{Alvarez}. It is well known that the isometric group is a Lie group in Riemannian manifold. This fact also holds in Finslerian manifold\cite{Deng}. The counterparts of Poincare algebra in Finsler spacetime could be studied. At last, we should give the kinematic and dispersion law in Finsler spacetime.

This paper is organized as follows. In Sec.2, we present basic notations of Finsler geometry and discuss inertial frames in Finsler spacetime. In Sec.3, we use the isometric group to investigate the symmetry of Finsler spacetime. In Sec.4, we discuss the kinematics in projectively flat $(\alpha,\beta)$ spacetime with constant flag curvature. The isometric groups and the corresponded Lie algebras for different types of $(\alpha,\beta)$ spacetime are given. At last, we give the concluding remarks. The counterpart of sectional curvature in Riemann geometry-flag curvature is introduced in appendix.

\section{Finsler spacetime}
The inertial frame means a particle in it continue at rest or in uniform straight motion. In an inertial system, the inertial motion is described by
\begin{equation}
\label{inertial}
x^i=v^i(t-t_0)+x^i_0,~~~v^i\equiv\frac{dx^i}{dt}=consts.
\end{equation}
It should be notice that such definition for inertial motion (\ref{inertial}) does not involve any specific requirements on the metric of spacetime. In fact, Einstein just assumed that the spacetime should be Euclidean which inherited from Newton\cite{Einstein}. If we loose the requirement that the spacetime should be Euclidean and require that the spacetime should be Riemannian, there exists three classes of inertial frames. Historically, de Sitter first used the projective coordinates (or Beltrami coordinates) to erect a spacetime with constant sectional curvature-the de Sitter spacetime. De Sitter used his dS spacetime to debate with Einstein on `relative inertial'. Actually, the dS spacetime is one kinds of locally projectively flat spacetime.

A spacetime is said to be locally projectively flat if at every point, the geodesics are straight lines
\begin{equation}
\label{defintion PF}
x^\mu(\tau)=f(\tau)m^\mu+n^\mu,
\end{equation}
where $\tau$ is the parameter of the curve, $f(\tau)$ is a function which depends on the metric of spacetime and $m^\mu, n^\mu$ are constants. Clearly, the definition of projectively flat spacetime (\ref{defintion PF}) implies the inertial motion. If $x^0$ denotes time, one could obtain the formula (\ref{inertial}) from (\ref{defintion PF}). In Riemannian manifold, Beltrami's theorem tells us that a Riemannian metric is locally projectively flat if and only if it is of constant sectional curvature. It is well known that there are three kinds of spacetime with constant sectional curvature. They are Minkowski (Mink) spacetime and dS/AdS spacetime. That is why there only exists three classes of inertial frames in Riemannian spacetime. The three classes of inertial frames are the basis of the dSSR.

If we further loose the requirement for spacetime, just require that the spacetime should be Finslerian, various inertial frames could be obtained, including the inertial frames for VSR and DSR.

Instead of defining an inner product structure over the tangent bundle in Riemann geometry, Finsler geometry is based on
the so called Finsler structure $F$ with the property
$F(x,\lambda y)=\lambda F(x,y)$ for all $\lambda>0$, where $x\in M$ represents position
and $y\equiv\frac{dx}{d\tau}$ represents velocity. The Finsler metric is given as\cite{Book
by Bao}
 \begin{equation}
 g_{\mu\nu}\equiv\frac{\partial}{\partial
y^\mu}\frac{\partial}{\partial y^\nu}\left(\frac{1}{2}F^2\right).
\end{equation}
Finsler geometry has its genesis in integrals of the form
\begin{equation}
\label{integral length}
\int^r_sF(x^1,\cdots,x^n;\frac{dx^1}{d\tau},\cdots,\frac{dx^n}{d\tau})d\tau~.
\end{equation}
The Finsler structure represents the length element of Finsler space. Two types of Finsler space should be noticed. One is the Riemann space. A Finsler metric is said to be Riemannian, if $F^2$ is quadratic in $y$. Another is locally Minkowski space. A Finsler metric is said to be locally Minkowskian if at every point, there is a local coordinate system, such that $F=F(y)$ is independent of the position $x$ \cite{Book by Bao}.

The geodesic equation for Finsler manifold is given as\cite{Book by Bao}
\begin{equation}
\label{geodesic}
\frac{d^2x^\mu}{d\tau^2}+2G^\mu=0,
\end{equation}
where
\begin{equation}
\label{geodesic spray}
G^\mu=\frac{1}{4}g^{\mu\nu}\left(\frac{\partial^2 F^2}{\partial x^\lambda \partial y^\nu}y^\lambda-\frac{\partial F^2}{\partial x^\nu}\right)
\end{equation} is called geodesic spray coefficient.
Obviously, if $F$ is Riemannian metric, then
\begin{equation}
G^\mu=\frac{1}{2}\gamma^\mu_{\nu\lambda}y^\nu y^\lambda,
\end{equation}
where $\gamma^\mu_{\nu\lambda}$ is the Riemannian Christoffel symbol. By making use of the geodesic equation (\ref{geodesic}), one could find that a Finsler metric is locally projectively flat if and only if $G^\mu$ satisfies
\begin{equation}
G^\mu=P(x,y)y^\mu,
\end{equation}
where $P(x,y)$ is a function of $x$ and $y$. It is equivalent to the following equation that was proposed by Hamel \cite{Hamel}
\begin{equation}
\label{Hamel}
\frac{\partial^2 F}{\partial x^\lambda \partial y^\nu}y^\lambda=\frac{\partial F}{\partial x^\nu}.
\end{equation}
By making use of the Hamel equation (\ref{Hamel}), we get
\begin{equation}
G^\mu=\left(\frac{\partial F}{\partial x^\nu} y^\nu/2F\right)y^\mu.
\end{equation}
It means that $P=\frac{\partial F}{\partial x^\nu} y^\nu/2F$. One should notice that
\begin{eqnarray}
\frac{dF}{d\tau}&=&\frac{\partial F}{\partial x^\mu}\frac{dx^\mu}{d\tau}+\frac{\partial F}{\partial y^\mu}\frac{dy^\mu}{d\tau}\nonumber\\
&=&2PF-2PF=0,
\end{eqnarray}
where we has already used the formula for $P$ and the geodesic equation (\ref{geodesic}) to deduce the second equation.

\section{Symmetry in Finsler space}
To investigate the Killing vectors, we should construct the isometric
transformations of Finsler structure. It is convenient to
discuss the isometric transformations under an infinitesimal
coordinate transformation for $x$
\begin{equation}
\label{coordinate tran}
\bar{x}^\mu=x^\mu+\epsilon V^\mu,
\end{equation}
together with a corresponding transformation for $y$
\begin{equation}
\label{coordinate tran1}
\bar{y}^\mu=y^\mu+\epsilon\frac{\partial V^\mu}{\partial x^\nu}y^\nu,
\end{equation}
where $|\epsilon|\ll1$.
Under the coordinate transformation (\ref{coordinate tran}) and (\ref{coordinate tran1}), to first order in $|\epsilon|$, we obtain the expansion of the Finsler structure,
\begin{equation}
\label{coordinate tran F}
\bar{F}(\bar{x},\bar{y})=\bar{F}(x,y)+\epsilon V^\mu\frac{\partial F}{\partial x^\mu}+\epsilon y^\nu\frac{\partial V^\mu}{\partial x^\nu}\frac{\partial F}{\partial y^\mu},
\end{equation}
where $\bar{F}(\bar{x},\bar{y})$ should equal to $F(x,y)$.
Under the transformation (\ref{coordinate tran}) and (\ref{coordinate tran1}), a Finsler structure is called isometry if and only if
\begin{equation}
F(x,y)=\bar{F}(x,y).
\end{equation}
Deducing from the (\ref{coordinate tran F}), we obtain the Killing equation $K_V(F)$ in Finsler space
\begin{equation}
\label{killing F}
K_V(F)\equiv V^\mu\frac{\partial F}{\partial x^\mu}+y^\nu\frac{\partial V^\mu}{\partial x^\nu}\frac{\partial F}{\partial y^\mu}=0.
\end{equation}

Searching the Killing vectors for general Finsler manifold is a difficult task. Here, we give the Killing vectors for a class of Finsler space-$(\alpha,\beta)$ space\cite{Shen} with metric defining as
\begin{eqnarray}
&&F=\alpha\phi(s),~~~s=\frac{\beta}{\alpha},\\
&&\alpha=\sqrt{a_{\mu\nu}y^\mu y^\nu}~~{\rm and}~~ \beta=b_\mu(x)y^\mu,
\end{eqnarray}
where $\phi(s)$ is a smooth function, $\alpha$ is a Riemannian metric and $\beta$ is a one form.
Then, the Killing equation (\ref{killing F}) in $(\alpha,\beta)$ space reads
\begin{eqnarray}
0&=&K_V(\alpha)\phi(s)+\alpha K_V(\phi(s))\nonumber\\\label{killing ori}
 &=&\left(\phi(s)-s\frac{\partial \phi(s)}{\partial s}\right)K_V(\alpha)+\frac{\partial\phi(s)}{\partial s}K_V(\beta).
\end{eqnarray}
And by making use of the Killing equation (\ref{killing F}), we obtain
\begin{eqnarray}
K_V(\alpha)&=&\frac{1}{2\alpha}(V_{\mu|\nu}+V_{\nu|\mu})y^\mu y^\nu,\\
K_V(\beta)&=&\left(V^\mu\frac{\partial b_\nu}{\partial x^\mu}+b_\mu\frac{\partial V^\mu}{\partial x^\nu}\right)y^\nu,
\end{eqnarray}
where $``|"$ denotes the covariant derivative with respect to the Riemannian metric $\alpha$.
The solutions of the Killing equation (\ref{killing ori}) have three viable scenarios. The first one is
\begin{equation}
\phi(s)-s\frac{\partial \phi(s)}{\partial s}=0~~{\rm and}~~ K_V(\beta)=0,
\end{equation}
which implies $F=\lambda\beta$ for all $\lambda\in\mathbb{R}$. The second one is
\begin{equation}
\frac{\partial\phi(s)}{\partial s}=0~~{\rm and}~~K_V(\alpha)=0,
\end{equation}
which implies $F=\lambda\alpha$ for all $\lambda\in\mathbb{R}$. The above two scenarios are just trivial space. Here we focus on the case of $\phi(s)-s\frac{\partial \phi(s)}{\partial s}\neq0$ and $\frac{\partial\phi(s)}{\partial s}\neq0$. This will induce the last scenario.

Apparently, in the last scenario we have such solutions
\begin{eqnarray}
\label{killing F1}
V_{\mu|\nu}+V_{\nu|\mu}&=&0,\\
\label{killing F2}
V^\mu\frac{\partial b_\nu}{\partial x^\mu}+b_\mu\frac{\partial V^\mu}{\partial x^\nu}&=&0.
\end{eqnarray}
The equation (\ref{killing F1}) is no other than the Riemannian Killing equation. The equation (\ref{killing F2}) can be regarded as the constraint on the Killing vectors that satisfy the Killing equation (\ref{killing F1}). Here, we must point out that additional solutions of Killing equation (\ref{killing ori}) for $(\alpha,\beta)$ space exist, besides the solutions (\ref{killing F1}) and (\ref{killing F2}). It will be discussed in next section.

However, the Killing equation for one type of $(\alpha,\beta)$ space-Randers space\cite{Randers} only have solutions (\ref{killing F1}) and (\ref{killing F2}). In Randers space, the $\phi(s)$ is set as $\phi(s)=1+s$. Then, the Killing equation (\ref{killing ori}) reduces to
\begin{equation}
\label{Randers killing}
K_V(\alpha)+K_V(\beta)=0.
\end{equation}
The $K_V(\alpha)$ contains irrational term of $y^\mu$ and $K_V(\beta)$ only contains rational term of $y^\mu$, therefore the equation (\ref{Randers killing}) satisfies if and only if $K_V(\alpha)=0$ and $K_V(\beta)=0$.

\section{Lie algebra and kinematics in projectively flat $(\alpha,\beta)$ spacetime}

An $n$ $(n>3)$ dimensional $(\alpha,\beta)$ space is projectively flat with constant flag curvature if and only if one of the following holds\cite{BLi}\\
\begin{itemize}
\item[A.] it is Riemann spacetime with constant sectional curvature;
\item[B.] it is locally Minkowski spacetime;
\item[C.] it is locally isometric to a generalized Funk spacetime\cite{Funk};
\item[D.] it is locally isometric to Berwald's metric spacetime\cite{Berwald}.
\end{itemize}
We will discuss the four types of projectively flat space respectively. Throughout this section the $\cdot$ denotes the inner product of Minkowski space $x\cdot x=\eta_{\mu\nu}x^\mu x^\nu$, where $\eta_{\mu\nu}={\rm diag}(1,-1,-1,-1)$.

\subsection{Symmetry in type A $(\alpha,\beta)$ spacetime and dSSR}
The metric of Riemann spacetime with constant sectional curvature can be given by the projective coordinate system
\begin{equation}
\label{Riem PF}
F_R=\frac{\sqrt{(y\cdot y)(1-\mu(x\cdot x))+\mu(x\cdot y)^2}}{1-\mu(x\cdot x)},
\end{equation}
where the sectional curvature $\mu$ of metric (\ref{Riem PF}) is constant. Clearly, the signature $+,0,-$ of $\mu$ corresponds to the dS spacetime, Mink spacetime and AdS spacetime, respectively. Such a metric (\ref{Riem PF}) is invariant under the fractional linear transformations (FLT), and it is $ISO(1,3)/SO(1,4)/SO(2,3)$- invariant Mink/dS/AdS-spacetime\cite{Guo}.

By making use of the formula (\ref{geodesic spray}), we know that the geodesic spray coefficient $G^\mu$ for metric (\ref{Riem PF}) is given as
\begin{equation}
G^\mu_R=\frac{\mu(x\cdot y)}{1-\mu(x\cdot x)}y^\mu.
\end{equation}
Thus, the geodesic equation for metric (\ref{Riem PF}) is of the form
\begin{equation}
\label{geodesic Riem}
\frac{d^2x^\mu}{d\tau^2}+\frac{2\mu(x\cdot \frac{dx}{d\tau})}{1-\mu(x\cdot x)}\frac{dx^\mu}{d\tau}=0.
\end{equation}
In fact, the geodesic equation is equivalent to
\begin{equation}
\label{P Riem}
\frac{dp^\mu}{d\tau}=0,~~p^\mu\equiv \frac{m_R}{F_R}\frac{1}{1-\mu(x\cdot x)}\frac{dx^\mu}{d\tau},
\end{equation}
where $m_R$ is the mass of the particle. Thus, $p^\mu$ is a constant along the geodesic. It could be regarded as the counterpart of momentum. 
From $F^2_R=g_{\mu\nu}y^\mu y^\nu$, we get
\begin{equation}
g_{\mu\nu}p^\mu p^\nu=\frac{1}{(1-\mu(x\cdot x))^2}m^2_R.
\end{equation}
It is obvious that if $\mu=0$, the above relation returns to the dispersion relation in Minkowski spacetime.
 The counterpart of angular momentum tensor could be defined as
\begin{equation}
L^{\mu\nu}\equiv x^\mu p^\nu-x^\nu p^\mu.
\end{equation}
It is also a conserved quantities along the geodesic, for $\frac{dL^{\mu\nu}}{d\tau}=0$.
The dispersion law in dSSR \cite{Guo} is given as
\begin{equation}
\label{dsr Riem}
p\cdot p-\frac{|\mu|}{2} L\cdot L=m_R^2.
\end{equation}

By making use of the Killing equation (\ref{killing F}), we obtain the Killing vectors for Riemmannian metric (\ref{Riem PF})
\begin{equation}
\label{KS Riem}
V^\mu=Q^\mu_{~\nu}x^\nu+C^\mu-\mu(x\cdot C)x^\mu,
\end{equation}
where $Q_{\mu\nu}=\eta_{\rho\mu}Q^\rho_{~\nu}$ is an arbitrary constant skew-symmetric matrix and $C_\mu=\eta_{\rho\mu}C^\rho$ is an arbitrary constant vector. The isometric group of a Finsler space is a Lie group \cite{Deng}. One should notice that translation-like generators are induced by $C^\mu$ and Lorentz generators are induced by $Q_{\mu\nu}$. The generators of isometric group in Riemannian space (\ref{Riem PF}) read
\begin{eqnarray}
\eta_{\mu\nu}\hat{p}^\nu=\hat{p}_\mu=i(\partial_\mu-\mu x_\mu(x\cdot\partial)),\\
\hat{L}_{\mu\nu}=x_\mu\hat{p}_\nu-x_\nu\hat{p}_\mu=i(x_\mu\partial_\nu-x_\nu\partial_\mu).
\end{eqnarray}
The non-trivial Lie algebra corresponded to the Killing vectors (\ref{KS Riem}) is given as
\begin{eqnarray}
~[\hat{p}_\mu,\hat{p}_\nu]&=&i\mu\hat{L}_{\mu\nu},\nonumber\\
\label{dSSR algebra}
~[\hat{L}_{\mu\nu},\hat{p}_\rho]&=&i\eta_{\nu\rho}\hat{p}_\mu-i\eta_{\mu\rho}\hat{p}_\nu,\\
~[\hat{L}_{\mu\nu},\hat{L}_{\rho\lambda}]&=&i\eta_{\mu\lambda}\hat{L}_{\nu\rho}-i\eta_{\mu\rho}\hat{L}_{\nu\lambda}+i\eta_{\nu\rho}\hat{L}_{\mu\lambda}-i\eta_{\nu\lambda}\hat{L}_{\mu\rho}\nonumber.
\end{eqnarray}
While sectional curvature of Riemannian spacetime (\ref{Riem PF}) $\mu$ vanishes, the dS/AdS spacetime reduce to Mink spacetime, the momentum tensors and angular momentum tensors reduce to the one in Mink spacetime, and the Lie algebra (\ref{dSSR algebra}) in dSSR reduces to Poincare algebra.

The sectional curvature $\mu$ is linked with \cite{Guo} the cosmological constant $\Lambda$ \cite{Copeland}, and the Newton-Hooke constant $\nu$ \cite{Huang}
\begin{equation}
\mu\simeq\frac{\Lambda}{3},~~\nu\equiv c\sqrt{\mu}\sim10^{-35}s^{-2}~.
\end{equation}

\subsection{Symmetry in type B $(\alpha,\beta)$ spacetime and VSR}
A Finsler metric is said to be locally Minkowskian if at every point, there is a local coordinate system, such that $F=F(y)$ is independent of the position $x$. It is clear from the definition (\ref{geodesic spray}) that the geodesic spray coefficient $G^\mu$ vanishes in locally Minkowski space. Thus, the geodesic equation of locally Minkowshi space is of simply form
\begin{equation}
\frac{d^2x^\mu}{d\tau^2}=0.
\end{equation}
The momentum tensor $p^\mu=\frac{m_V}{F_V}\frac{dx^\mu}{d\tau}$ and angular momentum tensor $L^{\mu\nu}\equiv x^\mu p^\nu-x^\nu p^\mu$ are conserved quantities along the geodesic, for
\begin{equation}
\frac{dp^\mu}{d\tau}=0,~~\frac{dL^{\mu\nu}}{d\tau}=0.
\end{equation}

Besides the Minkowski space, locally Minkowski space still involve a various types of metric space. But not all of them has physical implication. Here, we just focus on the locally Minkowski space which is invariant under the VSR symmetric group.

The VSR preserves the law of energy-momentum
conservation\cite{Glashow}. It implies that the translation
invariance should be contained in the symmetries of the VSR. The
left symmetries of the VSR include four possible subgroups of
Lorentz group. We introduce the notation  $T_1=(K_x+J_y)/\sqrt{2}$
and $T_2=(K_y-J_x)/\sqrt{2}$ (the index $x,y,z$ denote the space coordinate), where $J$ and $K$ are the generators
of rotations
and boosts, respectively. The four subgroups of Lorentz group are given as\cite{Sheikh}:\\
{\it i})$T(2)$, the Abelian subgroup of the Lorentz group, generated by $T_1$ and $T_2$, with the structure:
\begin{equation}
[T_1,T_2]=0;
\end{equation}
{\it ii})$E(2)$, the group of two-dimensional Euclidean motion, generated by $T_1$, $T_2$ and $J_z$, with the structure:
\begin{equation}
[T_1,T_2]=0,~[J_z,T_1]=-iT_2,~[J_z,T_2]=iT_1;
\end{equation}
{\it iii})$HOM(2)$, the group of orientation-preserving similarity transformations, generated by $T_1$, $T_2$
and $K_z$, with the structure:
\begin{equation}
[T_1,T_2]=0,~[T_1,K_z]=iT_1,~[T_2,K_z]=iT_2;
\end{equation}
{\it iv})$SIM(2)$, the group isomorphic to the four-parametric similitude group, generated by $T_1$, $T_2$, $J_z$ and $K_z$, with the structure:
\begin{eqnarray}
~[T_1,T_2]=0,&~[T_1,K_z]=iT_1&,~[T_2,K_z]=iT_2,\nonumber\\
~[J_z,K_z]=0,&~[J_z,T_1]=-iT_2&,~[J_z,T_2]=iT_1.
\end{eqnarray}

We will show that there is a relation between the isometric group of
the Finsler structure\cite{Gibbons}
\begin{equation}
\label{vsr metric}
F_V=(\eta_{\mu\nu}y^\mu y^\nu)^{(1-n)/2}(b_\rho y^\rho)^n
\end{equation}
and  symmetries of the VSR. Here $n$ is an arbitrary constant,
$\eta_{\mu\nu}$ is Minkowskian metric and
$b_\rho=\eta_{\mu\rho}b^\mu$ is a constant vector. It is referred as
the VSR metric. By making use of the Killing equation (\ref{killing
F}), we obtain Killing equation for the VSR metric
\begin{equation}
\label{killing vsr}
y^\nu\frac{\partial V^\mu}{\partial x^\nu}\left(\frac{(1-n)y_\mu(b_\rho y^\rho)^n+n(\eta_{\alpha\beta}y^\alpha y^\beta)^{1/2}b_\mu(b_\rho y^\rho)^{n-1}}{(\eta_{\alpha\beta}y^\alpha y^\beta)^{(1+n)/2}}\right)=0.
\end{equation}
The Eq. (\ref{killing vsr}) has solutions
\begin{eqnarray}\label{solu1}
V^{\mu}&=&Q^\mu_{~\nu} x^\nu+C^\mu,\\\label{solu2}
b_\mu Q^\mu_{~\nu}&=&0,
\end{eqnarray}
where $Q_{\mu\nu}=\eta_{\rho\mu}Q^\rho_{~\nu}$ is an arbitrary constant skew-symmetric matrix and $C_\mu=\eta_{\rho\mu}C^\rho$ is an arbitrary constant vector. If one requires that the transformation group for
the vectors no other than the Lorentz one or subgroup of Lorentz
one, formula (\ref{solu1}) togethers with the constraint
(\ref{solu2}) is the only solution of Killing equation (\ref{killing
F}) for the VSR metric.

Taking the light cone coordinate \cite{Kogut}
$\eta_{\alpha\beta}y^\alpha y^\beta=2y^+ y^--y^i y^i$ (with $i$
ranging over the values 1 and 2) and supposing
$b_\mu=\{0,0,0,b_-\}$($b_-=1$), we know that in general
$Q^{-}_{~\mu}\neq0$. It means that the Killing vectors of the VSR
metric (\ref{vsr metric}) do not have non-trivial components
$Q_{+-}$ and $Q_{+i}$. The isometric group of a Finsler space is a
Lie group \cite{Deng}. The non-trivial Lie algebra corresponded to
the Killing vectors (\ref{solu1}), which satisfies the constraint
(\ref{solu2}), is given as
\begin{eqnarray}
~[J_z,T^i]=i\epsilon_{ij}T^j,&~&[J_z,P^i]=i\epsilon_{ij}P^j,\nonumber\\
\label{E(2)} ~[T_i,P^-]=-iP_i,&~&[T_i,P^j]=-i\delta_{ij}P^+,
\end{eqnarray}
where $\epsilon_{12}=-\epsilon_{12}=1,
\epsilon_{11}=\epsilon_{22}=0$ and $P^\pm=(P_0\pm P_z)/\sqrt{2}$. It
is obvious that the generators of the isometric group of the VSR
metric include generators of $E(2)$  and  four spacetime translation
generators. This result induces the $E(2)$ scenario of VSR from the
VSR metric (\ref{vsr metric}). The $HOM(2)$ scenario of VSR could be
induced in the same approach.

The above investigations are under the premise that the direction of
spacetime is arbitrary or the transformation group for the vectors
no other than the Lorentz group or subgroups of Lorentz group. It
means that no preferred direction exists in spacetime. If the
spacetime does have a special direction, the Killing equation
(\ref{killing F}) for the VSR metric will have a special solution.
The VSR metric was first suggested by Bogoslovsky \cite{Bogoslovsky}.
He assumed that the spacetime has a preferred direction. Following
the assumption and taking the null direction to be the preferred
direction, we obtain the solution of Killing equation (\ref{killing
vsr})
\begin{equation}
\label{solu sim(2)}
V^\mu=(Q^\mu_{~\nu}+\delta^\mu_{~\nu}) x^\nu+C^\mu,
\end{equation}
where $Q^\mu_{~\nu}$ is  an antisymmetrical matrix and  satisfies
the requirement
\begin{equation}
Q_{+-}n^-=-n^-.
\end{equation}
Here $n^-$ is a null direction. One can check that the Killing
vectors (\ref{solu sim(2)}) does not have non-trivial components
$Q_{+i}$. It implies that the null direction is invariant under the
transformation
\begin{equation}
\Lambda^-_{~-}n^-\equiv\left(\delta^-_{~-}+\epsilon(n\delta^-_{~-}+Q^-_{~-})\right)n^-=\left(1+\epsilon(n-1)\right)n^-.
\end{equation}
Here, $\Lambda^\mu_{~\nu}$ denotes the counterpart of Lorentz
transformation. Therefore, if the spacetime has a preferred
direction in null direction, the symmetry corresponded to $Q_{+-}$
is restored. One can see that the Killing vectors (\ref{solu
sim(2)}) have a non-trivial component $\delta^\mu_{~\nu}x^\nu$. It
represents the dilations. Thus, we know that the transformation
group for the VSR metric (\ref{vsr metric}) contains dilations,
while the null direction is a  preferred direction. One could obtain
the Lie algebra for such transformation group. In fact, the
non-trivial Lie algebra is just the algebra of $DISIM(2)$ group
proposed by Gibbons {\it et al}.\cite{Gibbons}
\begin{eqnarray}
~[K_z,P^\pm]=-i(n\pm1)P^\pm,&~&[K_z,P^i]=-inP^i,\nonumber\\
~[K_z,T_i]=-iT_i,&~&[J_z,T^i]=i\epsilon_{ij}T^j,\nonumber\\
~[J_z,P^i]=i\epsilon_{ij}P^j,&~&[T_i,P^-]=-iP_i,\nonumber\\
\label{DISIM(2)} &~&[T_i,P^j]=-i\delta_{ij}P^+.
\end{eqnarray}
The $DISIM(2)$ group is a subgroup of Weyl group, it contains a
subgroup $E(2)$  together with a combination of a boost in the $+-$
direction and a dilation. It should be noticed that the deformed
generator $K_z$ acts not only as a boost but also a dilation. The
transformation acts by $K_z$ is given as
\begin{equation}
\bar{x}^\pm=\left(\exp(\phi)\right)^{\pm1+n}x^\pm, ~~\bar{x}^i=(\exp(\phi))^{n}x^i,
\end{equation}
where $\exp(\phi)=\sqrt{\frac{1+v/c}{1-v/c}}$. The transformations act by other generators of $DISIM(2)$ group are same with Lorentz one.

If $b_\mu$ in the  VSR metric (\ref{vsr metric}) has the form
$b_\mu=\{0,b_x,0,b_-\}$($b_x=b_-=1$), solutions of Killing equation
(\ref{killing vsr}) show that the Killing vectors just have
non-trivial components $Q_{-y}$ and $C^\mu$. However, the
corresponded Lie algebra does not exist. For the generators
corresponded to $Q_{-y}$ together with the generators of
translations can not form a subalgebra of the Poincare algebra.
Consequently, we show that the investigation of Killing equation for
VSR metric (\ref{vsr metric}) could account for the $E(2)$, $HOM(2)$
and $SIM(2)$($DISIM(2)$) scenarios of the VSR.

The Lagrangian for VSR metric is given as
\begin{equation}
\label{Lagrangian}
\mathcal{L}=m_VF_V=m_V(\eta_{\mu\nu}y^\mu y^\nu)^{(1-n)/2}(b_\rho y^\rho)^n.
\end{equation}
The corresponding dispersion relation is of the form
\begin{equation}
\label{dsr vsr}
\eta^{\mu\nu}p_\mu p_\nu=m_V^2(1-n^2)\left(\frac{n^\rho p_\rho}{m(1-n)}\right)^{2n/(1+n)}.
\end{equation}
The dispersion relation (\ref{dsr vsr}) is not Lorentz-invariant, but
invariant under the transformations of $DISIM(2)$ group. Ref.
\cite{Bogoslovsky} showed that the ether-drift experiments gives a
constraint $|n|<10^{-10}$ for the parameter $n$ of the VSR metric
(\ref{vsr metric}). 

\subsection{Symmetry in type C $(\alpha,\beta)$ spacetime}
The generalized Funk metric \cite{Funk} has two geometrical parameters. For physical consideration and simplicity, as DSR, VSR and dSSR, only one geometrical parameter is needed. Therefore, we just investigate the Funk metric of this form
\begin{equation}
\label{Funk metric}
F_F=\frac{\sqrt{(y\cdot y)(1-\kappa^2(x\cdot x))+\kappa^2(x\cdot y)^2}-\kappa(x\cdot y)}{1-\kappa^2(x\cdot x)}.
\end{equation}
Apparently, the Funk metric (\ref{Funk metric}) is of Randers type,
\begin{eqnarray}
F_F=\alpha_F+\beta_F,~~
\alpha_F=\frac{\sqrt{(y\cdot y)(1-\kappa^2(x\cdot x))+\kappa^2(x\cdot y)^2}}{1-\kappa^2(x\cdot x)},~~\beta_F=\frac{-\kappa(x\cdot y)}{1-\kappa^2(x\cdot x)}.
\end{eqnarray}
As discussed in Sec.3, the Killing vectors of Funk metric of Randers type must satisfy both $K_V(\alpha)=0$ and $K_V(\beta)=0$, and it is the only solutions of the Killing equation (\ref{killing F}).
The solution of equation $K_V(\alpha)=0$ gives
\begin{equation}
\label{solu Funk1}
V^\mu=Q^\mu_{~\nu}x^\nu+C^\mu-\kappa^2(x\cdot C)x^\mu,
\end{equation}
where $Q_{\mu\nu}=\eta_{\rho\mu}Q^\rho_{~\nu}$ is an arbitrary constant skew-symmetric matrix and $C_\mu=\eta_{\rho\mu}C^\rho$ is an arbitrary constant vector.
And the solution of equation $K_V(\beta)=0$ gives
\begin{equation}
\label{solu Funk2}
\kappa C^\nu=0.
\end{equation}
The  solutions (\ref{solu Funk1}) and (\ref{solu Funk2}) imply that the Killing vectors of Funk metric (\ref{Funk metric}) is of the form
\begin{equation}
\label{KS Funk}
V^\mu=Q^\mu_{~\nu}x^\nu,
\end{equation}
if $\kappa\neq0$. While $\kappa=0$, the Funk metric (\ref{Funk metric}) reduces to Minkowski metric, the solutions (\ref{solu Funk1}) and (\ref{solu Funk2}) reduce to
\begin{equation}
V^\mu=Q^\mu_{~\nu}x^\nu+C^\mu,
\end{equation}
as expected. The non-trivial Lie algebra of non-trivial Funk spacetime (\ref{Funk metric}) ($\kappa\neq0$) corresponded to the Killing vectors (\ref{KS Funk}) is given as
\begin{equation}
[\hat{L}_{\mu\nu},\hat{L}_{\rho\lambda}]=i\eta_{\mu\lambda}\hat{L}_{\nu\rho}-i\eta_{\mu\rho}\hat{L}_{\nu\lambda}+i\eta_{\nu\rho}\hat{L}_{\mu\lambda}-i\eta_{\nu\lambda}\hat{L}_{\mu\rho},
\end{equation}
where $\hat{L}_{\mu\nu}=i(x_\mu\partial_\nu-x_\nu\partial_\mu)$.
It means that the non-trivial Funk metric (\ref{Funk metric}) is invariant just under the Lorentz group.

By making use of the formula (\ref{geodesic spray}), the geodesic spray coefficient $G^\mu$ for metric (\ref{Funk metric}) is given as
\begin{equation}
G^\mu_F=-\kappa\frac{F_F}{2}y^\mu.
\end{equation}
Thus, the geodesic equation for metric (\ref{Funk metric}) is given as
\begin{equation}
\label{geodesic Funk}
\frac{d^2x^\mu}{d\tau^2}-\kappa F_F\frac{dx^\mu}{d\tau}=0.
\end{equation}
Actually, the geodesic equation (\ref{geodesic Funk}) is related to the scaled Berwald's metric $F_B$, which will be discussed in the next subsection. And the geometrical parameter in $F_B$ is set as $\kappa$. The derivative of $F_B$ with respect to the curve parameter $\tau$ in Funk metric (\ref{Funk metric}) reads
\begin{eqnarray}
\frac{dF_B}{d\tau}&=&\frac{\partial F_B}{\partial x^\mu}\frac{dx^\mu}{d\tau}+\frac{\partial F_B}{\partial y^\mu}\frac{dy^\mu}{d\tau}\nonumber\\
&=&-2\kappa F_B F_F+\kappa F_B F_F\nonumber\\
&=&-\kappa F_B F_F.
\end{eqnarray}
Therefore, the geodesic equation (\ref{geodesic Funk}) is equivalent to
\begin{equation}
\frac{dp^\mu}{d\tau}=0,~~p^\mu\equiv \frac{m_FF_B}{F_F^2}\frac{dx^\mu}{d\tau},
\end{equation}
where $m_F$ is the mass of the particle in Funk spacetime.

The dispersion relation in Funk spacetime (\ref{Funk metric}) is given as
\begin{equation}
\label{dsr Funk}
F^2_F(x,p)=g_{\mu\nu}p^\mu p^\nu=m_F^2\frac{F_F^2(x,y)}{F_B^2(x,y)}.
\end{equation}
The flag curvature of Funk spacetime is $K_F=\frac{1}{4}\kappa^2$. As the discussion about dSSR, the constant flag curvature may
relate to new physical scale (like cosmological constant), and it is very tiny. Therefore, such counterpart of special
relativity-Funk special relativity also cannot be excluded by the experiments.
To first order in $\kappa$, we obtain the expansion of the dispersion relation (\ref{dsr Funk})
\begin{equation}
\label{dsr Funk1}
p\cdot p-2\kappa(x\cdot p)\sqrt{p\cdot p}=m^2_F.
\end{equation}
Such dispersion relation (\ref{dsr Funk1}) could be regarded as one type of modified dispersion law in DSR.

\subsection{Symmetry in type D $(\alpha,\beta)$ spacetime}
The metric constructed by Berwald \cite{Berwald} is of the form
\begin{equation}
F=\frac{\left(\sqrt{(y\cdot y)(1-x\cdot x)+(x\cdot y)^2}+(x\cdot y)\right)^2}{(1-(x\cdot x))^2\sqrt{(y\cdot y)(1-x\cdot x)+(x\cdot y)^2}}.
\end{equation}
It is projectively flat with constant flag curvature $K_B=0$. One important property of projective geometry shows that a projectively flat space is still projectively flat after a scaling on $x$. It can be proved by using the Hamel equation (\ref{Hamel}). Thus, the scaled Berwald's metric is given as
\begin{equation}
\label{Berwald metric}
F_B=\frac{\left(\sqrt{(y\cdot y)(1-\lambda^2(x\cdot x))+\lambda^2(x\cdot y)^2}-\lambda(x\cdot y)\right)^2}{(1-\lambda^2(x\cdot x))^2\sqrt{(y\cdot y)(1-\lambda^2(x\cdot x))+\lambda^2(x\cdot y)^2}},
\end{equation}
where $\lambda$ is a constant. The flag curvature of scaled Berwald's metric (\ref{Berwald metric}) is $K_B=0$.

Defining
\begin{equation}
\alpha_B=\frac{\sqrt{(y\cdot y)(1-\lambda^2(x\cdot x))+\lambda^2(x\cdot y)^2}}{1-\lambda^2(x\cdot x)},~~\beta_B=\frac{-\lambda(x\cdot y)}{1-\lambda^2(x\cdot x)},
\end{equation}
we have
\begin{equation}
\label{Berwald metric1}
F_B=\frac{(\alpha_B+\alpha_B)^2}{\alpha_B(1-x\cdot x)}.
\end{equation}
Substituting the metric (\ref{Berwald metric1}) into the Killing equation (\ref{killing F}), we get
\begin{equation}
\label{Killing Berwald}
K_V(F_B)=\frac{\alpha_B+\beta_B}{\alpha_B(1-\lambda^2(x\cdot x))}\left((1-\beta_B/\alpha_B)K_V(\alpha_B)+2K_V(\beta_B)+2\lambda^2(\alpha_B+\beta_B)\frac{x_\mu V^\mu}{1-\lambda^2(x\cdot x)}\right)=0.
\end{equation}
The equations $K_V(\alpha_B)=0$ and $K_V(\beta_B)=0$ imply
\begin{equation}
\label{KS Berwald}
V^\mu=Q^\mu_{~\nu}x^\nu,
\end{equation}
if $\lambda\neq0$, where $Q_{\mu\nu}=\eta_{\rho\mu}Q^\rho_{~\nu}$ is an arbitrary constant skew-symmetric matrix. Furthermore, it is obvious that $x_\mu Q^\mu_{~\nu}x^\nu=0$. Therefore, the Killing vectors of the form (\ref{KS Berwald}) is a solution of the Killing equation (\ref{Killing Berwald}). The Killing vector of the form (\ref{KS Berwald}) means that the scaled Berwald's metric spacetime (\ref{Berwald metric}) is isotropic about a given point. Therefore, the Killing vectors which implies such symmetry (isotropic about a given point) reach its maximal numbers. And additional solutions of Killing equations (\ref{Killing Berwald}) must have the form
\begin{equation}
V^\mu=f^\mu(x,C),
\end{equation}
where $C_\mu=\eta_{\rho\mu}C^\rho$ is an arbitrary constant vector. If $V^\mu=\{f(x,c),0,0,0\}$ is a solution of Killing equation (\ref{Killing Berwald}), it is clear that $V^\mu=\{f(x,c),f(x,c),f(x,c),f(x,c)\}$ is also a solution of (\ref{Killing Berwald}). Therefore, the maximal dimension of isometric group of 4 dimensional scaled Berwald's spacetime equals either $6$ or $10$. It is known \cite{Egorov} that the maximal dimension of isometric group in an n dimensional non Riemannian Finslerian space is $\frac{n(n-1)}{2}+2$. The scaled Berwald's metric spacetime is non Riemannian. We conclude that the solution of Killing equation (\ref{Killing Berwald}) only have solutions of the form (\ref{KS Berwald}).

The Lie algebra of non-trivial scaled Berwald's metric spacetime (\ref{Berwald metric}) ($\lambda\neq0$) corresponded to the Killing vectors (\ref{KS Berwald}) is given as
\begin{equation}
[\hat{L}_{\mu\nu},\hat{L}_{\rho\lambda}]=i\eta_{\mu\lambda}\hat{L}_{\nu\rho}-i\eta_{\mu\rho}\hat{L}_{\nu\lambda}+i\eta_{\nu\rho}\hat{L}_{\mu\lambda}-i\eta_{\nu\lambda}\hat{L}_{\mu\rho},
\end{equation}
where $\hat{L}_{\mu\nu}=i(x_\mu\partial_\nu-x_\nu\partial_\mu)$.

By making use of the formula (\ref{geodesic spray}), we obtain the geodesic spray coefficient $G^\mu$ for metric (\ref{Berwald metric})
\begin{equation}
G^\mu_B=-\lambda\frac{\sqrt{(y\cdot y)(1-\lambda^2(x\cdot x))+\lambda^2(x\cdot y)^2}-\lambda(x\cdot y)}{1-\lambda^2(x\cdot x)}y^\mu=-\lambda F_Fy^\mu,
\end{equation}
where $F_F$ is the Funk metric, and the parameter in $F_F$ is set as $\lambda$.
Thus, the geodesic equation for metric (\ref{Berwald metric}) is given as
\begin{equation}
\label{geodesic Berwald}
\frac{d^2x^\mu}{d\tau^2}-2\lambda F_F\frac{dx^\mu}{d\tau}=0.
\end{equation}
One should notice that the derivatives of $F_F$ with respect to the curve parameter $\tau$ in scaled Berwald's metric (\ref{Berwald metric}) reads
\begin{equation}
\frac{dF_F}{d\tau}=\frac{\partial F_F}{\partial x^\mu}\frac{dx^\mu}{d\tau}+\frac{\partial F_F}{\partial y^\mu}\frac{dy^\mu}{d\tau}=-\lambda F_F^2+2\lambda F_F^2=-\lambda F_F^2.
\end{equation}
Therefore, the geodesic equation (\ref{geodesic Berwald}) is equivalent to
\begin{equation}
\frac{dp^\mu}{d\tau}=0,~~p^\mu\equiv \frac{m_BF_F^2}{F_B^3}\frac{dx^\mu}{d\tau},
\end{equation}
where $m_B$ is the mass of the particle in scaled Berwald's metric spacetime.
The dispersion relation in scaled Berwald's metric spacetime (\ref{Berwald metric}) is given as
\begin{equation}
\label{dsr Berwald}
F^2_B(x,p)=g_{\mu\nu}p^\mu p^\nu=m_B^2\frac{F_F^4(x,y)}{F_B^4(x,y)}.
\end{equation}
The parameter $\lambda$ in scaled Berwald's metric spacetime (\ref{Berwald metric}) may relate to new physical scale and it is very tiny.
To first order in $\lambda$, we obtain the expansion of the dispersion law (\ref{dsr Berwald})
\begin{equation}
\label{dsr Berwald1}
p\cdot p-4\lambda(x\cdot p)\sqrt{p\cdot p}=m^2_B.
\end{equation}

Here, we find that Funk spacetime (\ref{Funk metric}) and scaled Berwald's metric spacetime (\ref{Berwald metric}) have same isometric group. And the geodesic equations in Funk spacetime and scaled Berwald's metric spacetime are alike, if they both take the same geometrical parameter. Also, to first order in geometrical parameter, the dispersion relation are almost the same.

\section{Conclusion}
In this paper, we have extended the concept of inertial motion in the framework of the projective geometry. The inertial frames in projectively flat Finsler spacetime are investigated. We have studied the inertial motion in a special Finsler spacetime-the projectively flat $(\alpha,\beta)$ spacetime with constant flag curvature (the counterpart of sectional curvature). The projectively flat $(\alpha,\beta)$ spacetime with constant flag curvature can be divided into four types. We have showed that the inertial motion and symmetry in Type A and Type B spacetime are just the one in dSSR and VSR, respectively. And the dispersion law in Type C and Type D could be regarded as one types of modified dispersion law in DSR. The four types of $(\alpha,\beta)$ spacetime involve two parameters-the speed of light and a geometrical parameter which may relate to new physical scale. While the geometrical parameter vanishes, the four types of spacetime reduce to Minkowski spacetime, the momentum tensors and angular momentum tensors reduce to the one in Minkowski spacetime, the corresponded Lie algebra reduces to Poincare algebra, and the inertial motions reduce to the one in special relativity. In the following table, we list basic features of the kinematics and symmetry in the four types spacetime.

\begin{table}[htbp]\renewcommand{\arraystretch}{2.0}
\caption{the projectively flat $(\alpha,\beta)$ space with constant flag curvature}
\begin{center}
\begin{tabular}{|ccccc|}
\hline
Type~~~~&parameter~~~~~&geodesic equation~~~~~&momentum~~~~~&isometric group\\
A~~~~&$\mu$~~~~~~&$\frac{d^2x^\mu}{d\tau^2}+\frac{2\mu(x\cdot \frac{dx}{d\tau})}{1-\mu(x\cdot x)}\frac{dx^\mu}{d\tau}=0$~~~~~&$p^\mu_R\equiv m_R\frac{1}{F_R}\frac{1}{1-\mu(x\cdot x)}\frac{dx^\mu}{d\tau}$~~~~~&dS/AdS group\\
B~~~~&$n$~~~~~&$\frac{d^2x^\mu}{d\tau^2}=0$~~~~~&$p^\mu_V\equiv m_V\frac{1}{F_V}\frac{dx^\mu}{d\tau}$~~~~~&DISIM(2) group\\
C~~~~&$\kappa$~~~~~&$\frac{d^2x^\mu}{d\tau^2}-\kappa F_F(\kappa)\frac{dx^\mu}{d\tau}=0$~~~~~&$p^\mu_F\equiv m_F\frac{F_B(\kappa)}{F_F^2(\kappa)}\frac{dx^\mu}{d\tau}$~~~~~&Lorentz group\\
D~~~~&$\lambda$~~~~~&$\frac{d^2x^\mu}{d\tau^2}-2\lambda F_F(\lambda)\frac{dx^\mu}{d\tau}=0$~~~~~&$p^\mu_B\equiv m_B\frac{F_F^2(\lambda)}{F_B^3(\lambda)}\frac{dx^\mu}{d\tau}$~~~~~&Lorentz group\\
\hline
\end{tabular}
\end{center}
\end{table}

\vspace{1cm}
\begin{acknowledgments}
We would like to thank Prof. C. J. Zhu, C. G. Huang and Z. Shen for useful discussions. The
work was supported by the NSF of China under Grant No. 10525522,
10875129 and 11075166.

\appendix*
\section{Flag curvature}
The flag curvature \cite{Book by Bao,Shen1} in Finsler geometry is the counterpart of the sectional curvature in Riemannian geometry. It is a geometrical invariant. Furthermore, the same flag curvature is obtained for any connection chosen in Finsler space. The curvature tensor $R^\mu_{~\nu}$ is defined as
\begin{equation}
\label{predecessor flag}
R^\mu_{~\nu}(x,y)\equiv-\left(\frac{\partial G^\mu}{\partial x^\nu}-y^\lambda\frac{\partial^2 G^\mu}{\partial x^\lambda\partial y^\nu}+2G^\lambda\frac{\partial^2 G^\mu}{\partial y^\lambda\partial y^\nu}-\frac{\partial G^\mu}{\partial x^\lambda}\frac{\partial G^\lambda}{\partial x^\nu}\right),
\end{equation}
where $G^\mu$ is geodesic spray coefficient. For a tangent plane $\Pi\subset T_xM$ and a non-zero vector $y\in T_xM$, the flag curvature is defined as
\begin{equation}
\label{flag curvature}
K(\Pi,y)\equiv\frac{g_{\lambda\mu}R^\mu_{~\nu}u^\nu u^\lambda}{F^2g_{\rho\theta}u^\rho u^\theta-(g_{\sigma\kappa}y^\sigma u^\kappa)^2},
\end{equation}
where $u\in\Pi$. If $F$ is projectively flat, substituting $G^\mu=P(x,y)y^\mu$ into the definition of flag curvature (\ref{flag curvature}), and by making use of formula (\ref{predecessor flag}), we obtain that
\begin{equation}
\label{flag curvature1}
K=-\frac{P^2-\frac{\partial P}{\partial x^\mu}y^\mu}{F^2}.
\end{equation}
The curvature tensor $R^\mu_{~\nu}$ defined above is presented as $-\bar{R}^\mu_{~\nu}$ in Ref.\cite{Shen1}. The notation we used here keeps the sectional curvature of dS spacetime to be positive and of AdS spacetime to be negative.
By making use of the formula for the flag curvature of projectively flat Finsler spacetime (\ref{flag curvature1}), we get the flag curvature for dS/AdS spacetime (\ref{Riem PF}), Funk spacetime (\ref{Funk metric}) and scaled Berwald's metric spacetime (\ref{Berwald metric}), respectively,
\begin{equation}
K_R=\mu,~~~K_F=\frac{1}{4}\kappa^2,~~~K_B=0
\end{equation}
And the flag curvature of locally Minkowski spacetime equals zero.

\end{acknowledgments}

\end{document}